\documentclass[runningheads]{llncs}
\usepackage{graphicx}
\usepackage{multirow}
\usepackage{booktabs}
\usepackage{tabularx}
\usepackage{fixltx2e}
\usepackage{amsmath}
\usepackage{amssymb}
\begin{document}
\title{Over-and-Under Complete Convolutional RNN for MRI Reconstruction}
%
%\titlerunning{Overcomplete Representations for MRI Reconstruction Using CRNN}
\titlerunning{Over-and-Under Complete Convolutional RNN for MRI Reconstruction}
\author{Pengfei Guo\inst{1}, Jeya Maria Jose Valanarasu\inst{2}, Puyang Wang\inst{2}, Jinyuan Zhou\inst{3}, Shanshan Jiang\inst{3}, Vishal M. Patel\inst{1,2}}
\authorrunning{Guo et al.}
\institute{Department of Computer Science, Johns Hopkins University, MD, USA\\
	\and
	Department of Electrical and Computer Engineering, Johns Hopkins University, MD, USA\\
	\and
	Department of Radiology, Johns Hopkins University, Baltimore, MD, USA}
\maketitle              % typeset the header of the contribution
\begin{abstract}
Reconstructing magnetic resonance (MR) images from under-sampled data is a challenging 
problem due to various artifacts introduced by the under-sampling operation. Recent deep 
learning-based methods for MR image reconstruction usually leverage a generic auto-encoder 
architecture which captures low-level features at the initial layers and high-level features at the deeper layers.  Such networks focus much
on global features which may not be optimal to reconstruct the fully-sampled image. In this paper, we propose an \textbf{O}ver-and-\textbf{U}nder \textbf{C}omplete Convolutional \textbf{R}ecurrent Neural Network (OUCR), which consists of an overcomplete and an undercomplete Convolutional Recurrent Neural Network (CRNN). The overcomplete branch gives special attention in learning local structures by restraining the receptive field of the network. Combining it with the undercomplete branch leads to a network which focuses more on low-level features without losing out on the global structures. Extensive experiments on two datasets demonstrate that the proposed method achieves significant improvements over the compressed sensing and popular deep learning-based methods with less number of trainable parameters.

\keywords{Convolutional RNN  \and MRI reconstruction \and Deep Learning.}
\end{abstract}
\section{Introduction}
Magnetic resonance imaging (MRI) is a noninvasive medical imaging approach that provides various tissue contrast mechanisms for visualizing anatomical structures and functions. Due to the hardware constraint, one major limitation of MRI is relatively slow data
acquisition process, which subsequently causes higher imaging cost and patients' discomfort in many clinical applications~\cite{au1}. While the exploitation of advanced hardware and parallel imaging~\cite{au2} can mitigate such issue, a common approach is to shorten the image acquisition time by under-sampling k-space (also known as Compressed Sensing (CS)) \cite{au5,au34}. However, reconstructing an image directly from partial k-space data results in a suboptimal image with aliasing artifacts. To deal with this issue, nonlinear recovery algorithms based on $\ell_{0}$, $\ell_{1}$ or total variation minimization are often used to recover the image from incomplete k-space data. More advanced CS-based image reconstruction algorithms have been combined with  parallel imaging~\cite{au6}, low-rank constraint terms~\cite{au7}, and dictionary learning~\cite{au8}.  Unfortunately, though CS image reconstruction algorithms are able to recover images, they lack noise-like textures and as a result many physicians find CS reconstructed images as ``artificial".  Moreover, when large errors are not reduced during optimization, high-frequency oscillatory artifacts cannot be properly removed~\cite{au9}. Thus, the acceleration factors of CS-based algorithms are generally limited between 2.5 and 3 for typical MR images~\cite{au9}. 

Recent advances in deep neural networks open a new possibility to solve the inverse problem of MR image reconstruction in an efficient manner~\cite{au35}. Artificial neural network-based image reconstruction methods have been shown to provide much better MR image quality than conventional CS-based methods \cite{au11,au27,au12,au30,au10,Deep_MRI_SPM,au14,au13,au15}.  
%Various deep networks have been proposed in the literature which can reconstruct an image from under-sampled data in different domains~\cite{au11,au12,au10}. Some of the other advanced deep learning (DL) based methods include cascaded architectures~\cite{au13}, invertible models~\cite{au14}, neural ODEs~\cite{au27}, and convolutional recurrent neural network ( )~\cite{au15}.
Most DL-based methods are convolutional which consist of a set of convolution and down/up sampling layers for efficient learning. The main intuition behind this kind of network architecture is that
at the initial layers, the receptive field  of the filters is smaller, so low-level features (e.g., edges) are captured. At deeper layers, the receptive field of the filters is larger, so high-level features (e.g., the interpretation of input) are captured. Using such generic architecture for MR image reconstruction might not be optimal, since low-level vision tasks are mainly concerned with extracting descriptions from input rather than the interpretation of input~\cite{au16}. Prior to DL era,  overcomplete representations were explored for dealing with noisy observations in the vision and image processing tasks~\cite{au17,au18}. In an overcomplete architecture, the increase of receptive field is restricted through the network, which forces the filters to focus on low-level features~\cite{au19}. Recently, overcomplete representations have been explored for image segmentation~\cite{au20} and image restoration~\cite{au19}. 

In this paper, a novel \textbf{O}ver-and-\textbf{U}nder \textbf{C}omplete Convolutional \textbf{R}ecurrent Neural Network (OUCR) is proposed that can recover a fully-sampled image from the under-sampled k-space data for accelerated MR image reconstruction. To summarize, the following are our key contributions: \textbf{1.} An over-complete convolutional RNN architecture is explored for MR image reconstruction. \textbf{2.} To recover finer details better, the proposed OUCR consists of two branches that can leverage the features of both undercomplete and overcomplete CRNN. \textbf{3.} Extensive experiments are conducted on two datasets and it is demonstrated that the proposed method achieves significant improvements over CS-based as well as popular DL-based methods and is more parameter efficient.

\section{Methodology}
\begin{figure}[t]
	\centering
	\includegraphics[width=\textwidth]{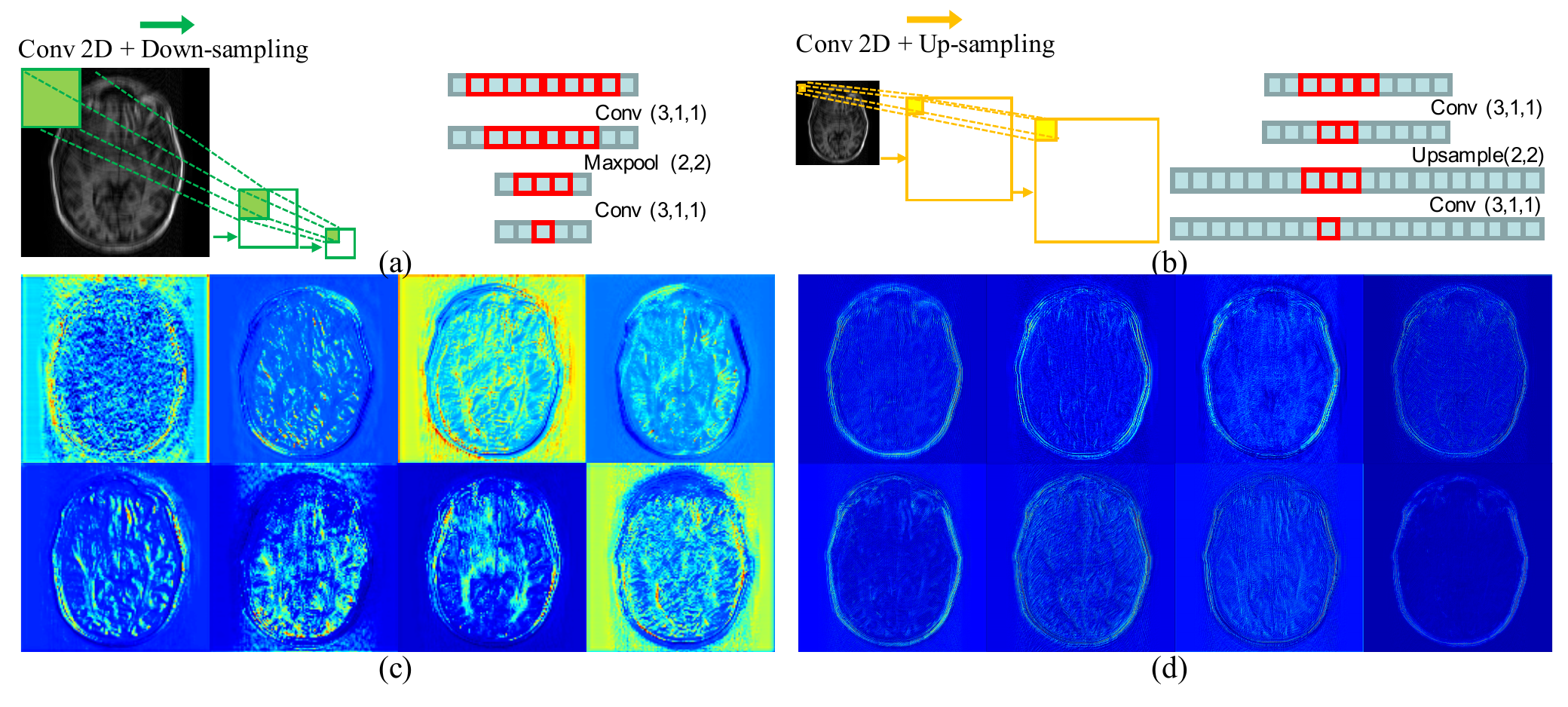}
	\vskip-13pt	
	\caption{\emph{Top row}: Explanation for the receptive field change in (a) undercomplete and (b) overcomplete networks. Conv(3,1,1) represents a convolution layer with kernel size as 3, stride as 1 and padding as 1. Upsample (2,2) represents a nearest neighbor upsampling layer with upsampling coefficients as (2,2). Maxpool(2,2) represents maxpooling layer with kernel size as 2 and stride as 2. The red pixels denote receptive field. It can be noted that the receptive field is constrained in overcomplete network compared to undercomplete network. \emph{Bottom row}: Visualization of filter responses (feature maps from ResBlock) for (c) undercomplete and (d) overcomplete CRNN. By restricting the size of receptive field, the overcomplete CRNN is able to focus on low-level features.} \label{fig1}
\end{figure}

\noindent {\bf{Overcomplete Networks. }} Overcomplete representations were first explored for signal representation where overcomplete bases were used such that the number of basis functions are more than the input signal samples \cite{au17}. This enabled a high flexibility at expressing the structure of the data. In neural networks, overcomplete fully connected networks were observed to be better feature extractors and hence perform well at denoising \cite{au18}. Recently, overcomplete convolutional networks are noted to be better at extracting local features because of restraining the receptive field when compared to the generic encoder-decoder networks~\cite{au25,au33,au20,au19}. Rather than using max-pooling layers at the encoder like an undercomplete convolutional network, in an overcomplete convolutional network, one uses upsampling layers. Fig.~\ref{fig1}(a) and (b) visually explains this concept in CNNs. As can be seen from Fig.~\ref{fig1}(c) and (d), the overcomplete networks focus on fine structures and learning local features from an image as the receptive field is constrained even at deeper layers. More examples of comparison between over and under complete networks can be found in the supplementary material. 
   
%   \begin{figure}[t]
%   	\centering
%   	\includegraphics[width=0.7\textwidth]{fig/rf.png}
%   	\vskip-13pt	
%   	\caption{Receptive field size change in an (a) overcomplete network (b) undercomplete network. Conv(3,1,1) represents a convolution layer with kernel size as 3, stride as 1 and padding as 1. Up-sample (2,2) represents a bilinear upsampling layer with upsampling coefficients as (2,2). Maxpool(2,2) represents maxpooling layer with coefficient 2. The red pixels highlighted denote the receptive field. It can be noted that the receptive field is constrained in overcomplete network compared to undercomplete network.} \label{rf}
%   \end{figure}
   
\noindent {\bf{MR Image Reconstruction. }} Let $x\in \mathbb{C}^N$ denote the observed under-sampled k-space data, $y\in \mathbb{C}^M$ is the fully-sampled image that we want to reconstruct. To obtain a regularized solution, the optimization problem of MRI reconstruction can be formulated as follows~\cite{au21,au13}:
\setlength{\belowdisplayskip}{0pt} \setlength{\belowdisplayshortskip}{0pt}
\setlength{\abovedisplayskip}{0pt} \setlength{\abovedisplayshortskip}{0pt}
\begin{equation} \label{eq2}
	\underset{y}{\mathrm{min}}   R(y) + \lambda \|x - F_{D}y\|_2^2 .
\end{equation}
Here, $R(y)$ is the regularization term and $\lambda$ controls the contribution of second term. $F_{D}$ represents the undersampling Fourier encoding matrix that is defined as the multiplication of the Fourier
transform matrix with a binary undersampling mask $D$. The ratio of the amount of k-space data required for a fully-sampled image to the amount collected in an accelerated acquisition is controlled by the acceleration factor (AF). The approximate fully-sampled image $\bar{y}$ can be measured from the observed under-sampled k-space data $x$ via an optimization process. One can solve the objective function (Eq.~\ref{eq2}) based on iterative optimization methods, such as gradient descent. A CRNN~\cite{au21,au15} is capable of modeling the iterative optimization process in Eq.~\ref{eq2} as $\bar{y} = \text{CRNN}(\bar{x},x,D,\Theta),$
where $\bar{y}$ is the reconstructed MR image from CRNN model, $\bar{x}$ is the zero-filled image and $\Theta$ denotes the trainable parameters of the CRNN model. 
\begin{figure}[t]
	\centering
	\includegraphics[width=\textwidth]{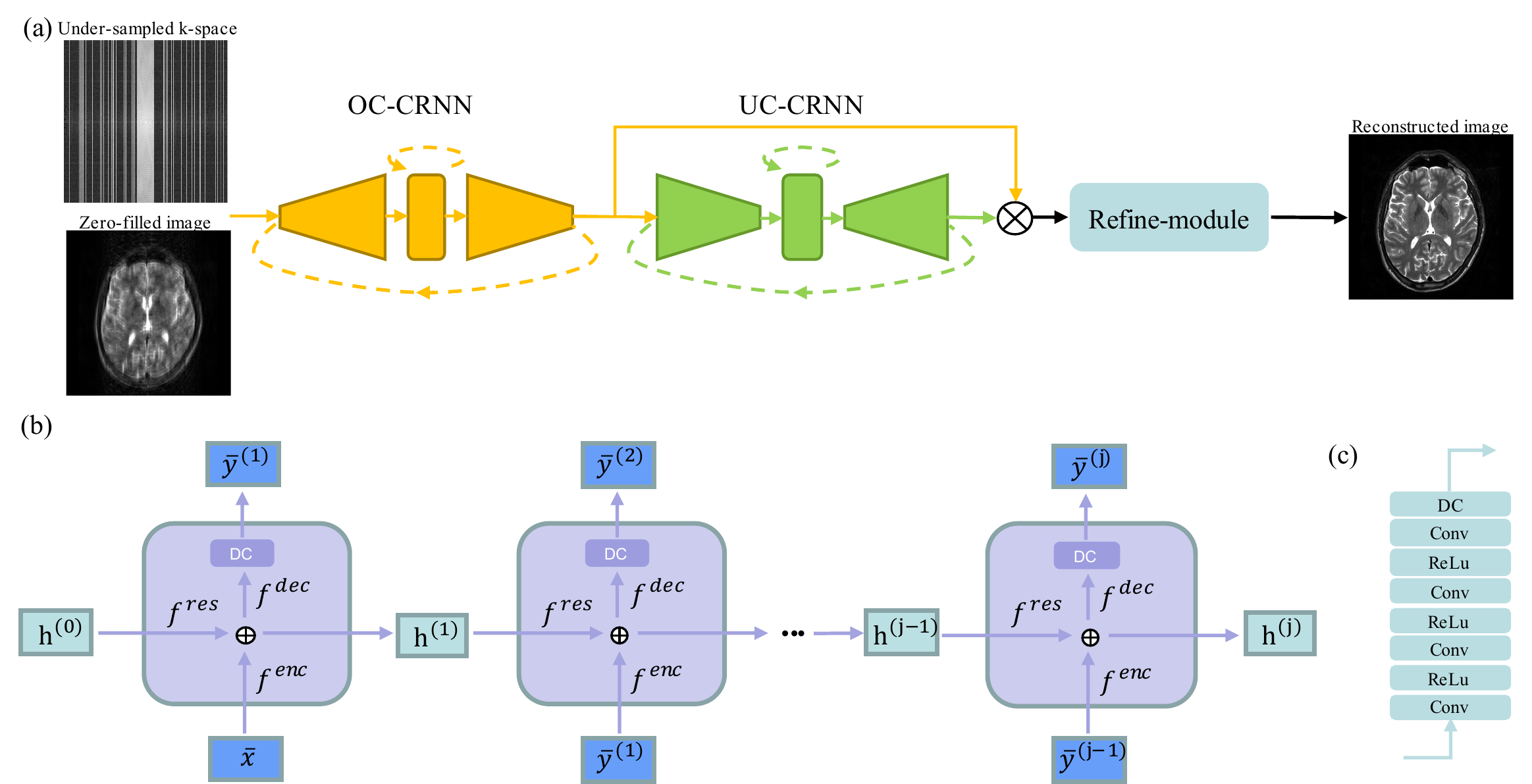}
	\vskip-12pt	\caption{(a) An overview of the proposed OUCR. Here, $\otimes$ denotes the channel-wise concatenation. We denote overcomplete and undercomplete CRNN as OC-CRNN and UC-CRNN, respectively. (b) A schematic of unrolled CRNN iterations. Here, $\oplus$ denotes the element-wise addition. (c) The network configuration of the refine-module. DC represents the data consistency layer.  \label{fig2} }
	\vskip-18pt
\end{figure}

\noindent {\bf{OUCR. }} To have special attention in learning low-level feature structures while not losing out on the global structures and inspired by previous CRNN methods~\cite{au21,au15},  we propose a novel OUCR network  as
shown in Fig.~\ref{fig2}. OUCR consists of two CRNN modules with different receptive fields to reconstruct MR images (i.e. OC-CRNN and UC-CRNN in Fig.\ref{fig2}(a)) and an refine-module (RM). In a CRNN module  $f_i$, let $f_i^{\text{enc}}$, $f_i^{\text{dec}}$, and $f_i^{\text{res}}$ denote the encoder, decoder, and ResBlock, respectively (Fig.\ref{fig2}(b)). A data consistency (DC) layer is added at the end of each module to reinforce the data consistency in k-space. The iterations of a CRNN module can be unrolled as follows:
\begin{equation} \label{eq5}
\begin{aligned} 
\bar{y}_i^{(j+1)} &= \text{DC}(f_i(\bar{y}_i^{(j)},h_i^{(j)}),x,D), \\
 &= F^{-1}[Dx+(1-D)F[f_i(\bar{y}_i^{(j)},h_i^{(j)})]],\\
 &= F^{-1}[Dx+(1-D)F[f_i^{\text{dec}}(f_i^{\text{res}}(h_i^{(j)}) + f_i^{\text{enc}}(\bar{y}_i^{(j)}) ) ]],\\
\end{aligned}
\end{equation}
where $h_i^{j}$ is the hidden state after iteration $j$ and $F^{-1}$ denotes the inverse Fourier transform. After $J$ iterations of the two CRNN modules, the final reconstructed MR image $\bar{y}$ is formulated as follows:
\begin{equation} \label{eq6}
\begin{aligned} 
\bar{y}_{\text{oc}} &= \text{OC-CRNN}(\bar{x},x,D,\Theta_{\text{oc}}), \\
\bar{y}_{\text{uc}} &= \text{UC-CRNN}(\bar{y}_{\text{oc}},x,D,\Theta_{\text{uc}}),\\
\bar{y} &= \text{DC}(\text{RM}(\bar{y}_{\text{oc}} \otimes \bar{y}_{\text{uc}}, \Theta_{\text{rm}}), x, D),\\
\end{aligned}
\end{equation}
where $\otimes$ denotes the channel-wise concatenation and $\Theta_{\text{oc}}, \Theta_{\text{uc}}, \Theta_{\text{rm}}$ denote  the  parameters of the overcomplete, undercomplete CRNN and RM network, respectively.
The intuition behind using both OC-CRNN and UC-CRNN is to make use of both local and global features. While
we focus more on the local features using the OC-CRNN, the global features are not neglected altogether as
they still have meaningful information for proper
reconstruction. In each CRNN module, we have
two convolutional blocks in both encoder and decoder. Each
convolutional block in the encoder has a 2D convolutional
layer followed by an upsampling layer in OC-CRNN or a max-pooling
layer in UC-CRNN. In the decoder, each convolutional
block has a 2D convolutional layer followed by max-pooling layer in OC-CRNN or upsampling layer in UC-CRNN.
%All the convolutional layers have a 3 $\times 3$  kernel, stride of 1 and padding of 1 unless mentioned otherwise. For upsampling, we perform nearest neighbor upsampling with a scale factor of 2. The max-pooling layer has a 2 $\times$ 2 kernel, stride of 2 and padding of 0. Each convolutional layer (except the last one) is followed by ReLU activation. 
More details regarding the network configuration can be found in the supplementary material.

\section{Experiments and Results}
\noindent {\bf{Evaluation and Implementation Details. }} The following two datasets are used for conducting experiments -- \textbf{fastMRI}~\cite{au22} and \textbf{HPKS}~\cite{au32,au31}.  The \textbf{fastMRI} dataset consists of single-coil coronal proton density-weighted knee images corresponding to 1172 subjects. In particular, 973 subjects' data is used for training, and 199 subjects' data (fastMRI validation dataset) is used for testing. For each subject, there are approximately 35 knee images that contain tissues. The \textbf{HPKS} dataset is collected by an anonymous medical center from post-treatment patients with malignant glioma. $T_1$-weighted images from 144 subjects are used, where 102 subjects' data are used for training, 14 subjects' data set are used for validation, and 28 subjects' data are used for testing. For each subject, 15 axial cross-sectional images that contain brain tissues are provided in this dataset. We simulated k-space measurements using the same sampling mask function as the fastMRI challenge~\cite{au22} with 4$\times$ and 8$\times$ accelerations. All models were trained using the $\ell_{1}$ loss with Adam optimizer by the following hyperparameters: initial learning rate of $1.5 \times 10^{-4}$ then reduced by a factor of 0.9 every 5 epochs; 50 maximum epochs; batch size of 4; the number of CRNN iteration $J$ of 5.  SSIM and PSNR are used as the evaluation metrics for comparison.

\noindent {\bf{MR Image Reconstruction Results. }} Table~\ref{tab1} shows the results corresponding to seven different methods evaluated on the HPKS and fastMRI datasets. The performance of the proposed model was compared with compressed sensing (CS)~\cite{au23}, UNet~\cite{au24},  KIKI-Net~\cite{au12}, Kiu-net~\cite{au20}, D5C5~\cite{au13}, and PC-RNN~\cite{au15}. For a fair comparison, UNet~\cite{au24} and Kiu-net~\cite{au20} are modified for data with real
and imaginary channels and a DC layer is added at the end of the networks. KIKI-Net~\cite{au12} which conducts interleaved convolution operation on image and k-space domains achieves better performance than UNet~\cite{au24} on HPKS. Kiu-net~\cite{au20} which is an overcomplete variant architecture of UNet~\cite{au24} outperforms UNet~\cite{au24} and KIKI-Net~\cite{au12}. By leveraging the cascade of convolutional neural networks, D5C5~\cite{au13} outperforms the Kiu-net~\cite{au20} in both HPKS and fastMRI datasets. PC-RNN~\cite{au15} which learns the
mapping in an iterative way by CRNN from three different scales achieves the second best performance. As it can be seen from Table~\ref{tab1}, the proposed OUCR outperforms other methods by leveraging the overcomplete architecture. Fig.~\ref{fig3} shows the qualitative results of two datasets with 4$\times$ and 8$\times$ accelerations. It can be observed that the proposed OUCR yields reconstructed images with remarkable visual similarity to the reference images compared to the others (see the last column of each sub-figure in Fig.~\ref{fig3}) in two datasets with different modalities.
The reported improvements achieved by OUCR are statistically significant ($p<10^{-5}$). The detailed statistical significance test is provided in Supplementary Table 1. 

\begin{table}[t!]
	\centering
	\setlength{\tabcolsep}{4.5pt}
	\scriptsize
	\caption{Quantitative results on the HPKS and fastMRI dataset. Param denotes the number of parameters.}\label{tab1}
	\vskip-12pt
\begin{tabular}{c|c|c|c|c|c|c|c|c|c}
	\hline
	\multirow{3}{*}{Method} & \multirow{3}{*}{Param} & \multicolumn{4}{c|}{HPKS}                                           & \multicolumn{4}{c}{fastMRI}                                        \\ \cline{3-10} 
	&                        & \multicolumn{2}{c|}{PSNR}       & \multicolumn{2}{c|}{SSIM}         & \multicolumn{2}{c|}{PSNR}       & \multicolumn{2}{c}{SSIM}         \\ \cline{3-10} 
	&                        & 4X             & 8X             & 4X              & 8X              & 4X             & 8X             & 4X              & 8X              \\ \hline
	CS                      & -                      & 29.94          & 24.96          & 0.8705          & 0.7125          & 29.54          & 26.99          & 0.5736          & 0.4870          \\ \hline
	UNet                    & 8.634 M                & 34.47          & 29.47          & 0.9155          & 0.8249          & 31.88          & 29.78          & 0.7142          & 0.6424          \\ \hline
	KIKI-Net                & 1.790 M                & 35.35          & 29.86          & 0.9363          & 0.8436          & 31.87          & 29.27          & 0.7172          & 0.6355          \\ \hline
	Kiu-net                 & 7.951 M                & 35.35          & 30.18          & 0.9335          & 0.8467          & 32.06          & 29.86          & 0.7228          & 0.6456          \\ \hline
	D5C5                    & 2.237 M                & 37.51          & 30.40          & 0.9595          & 0.8623          & 32.25          & 29.65          & 0.7256          & 0.6457          \\ \hline
	PC-RNN                  & 1.482 M                & 38.36          & 31.54          & 0.9696          & 0.8965          & 32.37          & 30.17          & 0.7281          & 0.6585          \\ \hline
	OUCR                    & \textbf{1.192 M}       & \textbf{39.33} & \textbf{32.14} & \textbf{0.9747} & \textbf{0.9044} & \textbf{32.61} & \textbf{30.59} & \textbf{0.7354} & \textbf{0.6634} \\ \hline
\end{tabular}
\vskip-18pt
\end{table}
\begin{figure}[]
	\centering
	\includegraphics[width=\textwidth]{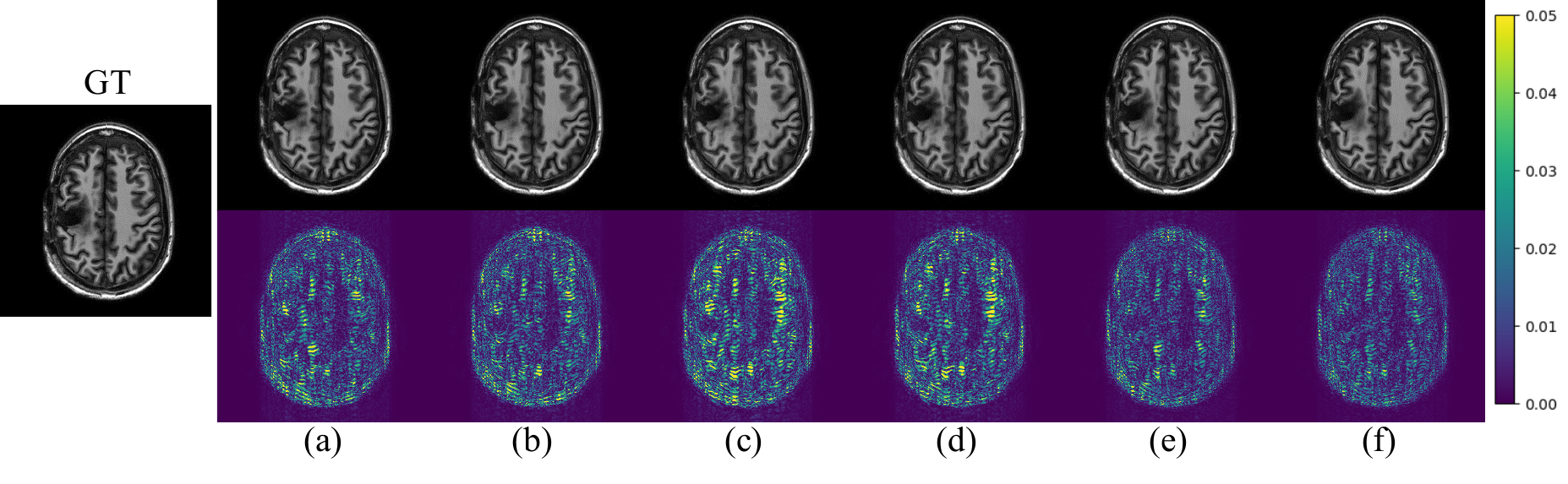}
	\vskip-12pt	\caption{Qualitative results and error maps of ablation study on
		HPKS with AF=4. (a) UC-CRNN. (b) UC-CRNN + RM. (c) OC-CRNN. (d) OC-CRNN + RM. (e) UC-CRNN + OC-CRNN. (f) UC-CRNN + OC-CRNN + RM (proposed OUCR). \label{fig5}}

\end{figure}
\begin{figure}[ht!]
	\centering
	\includegraphics[width=\textwidth]{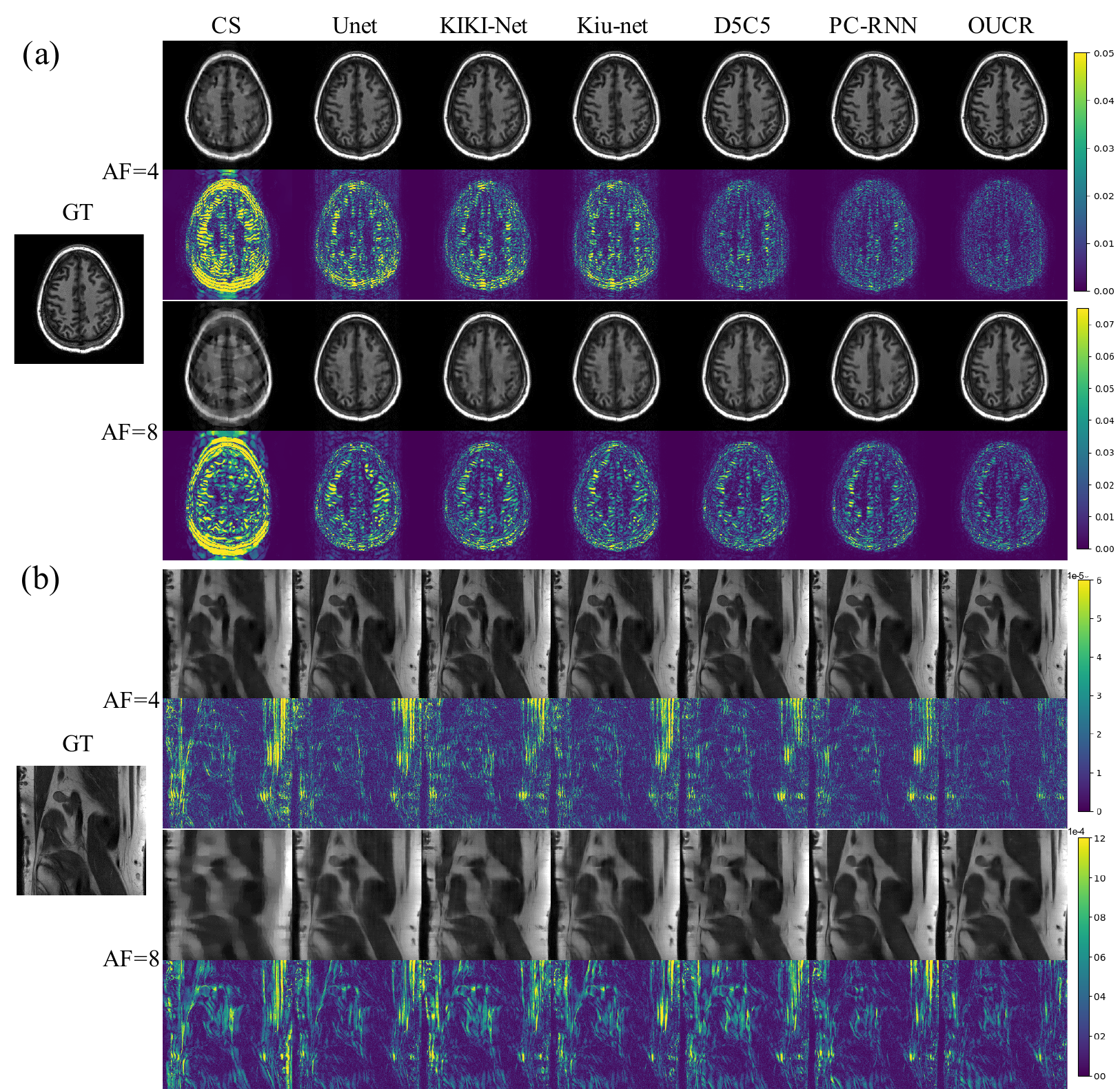}
	\vskip-12pt	\caption{Qualitative comparison of different methods on (a) HPKS and (b) fastMRI dataset. The second row of each subplot shows the corresponding error maps. \label{fig3}}
\end{figure}

\begin{figure}[htp!]
	\centering
	\includegraphics[width=\textwidth]{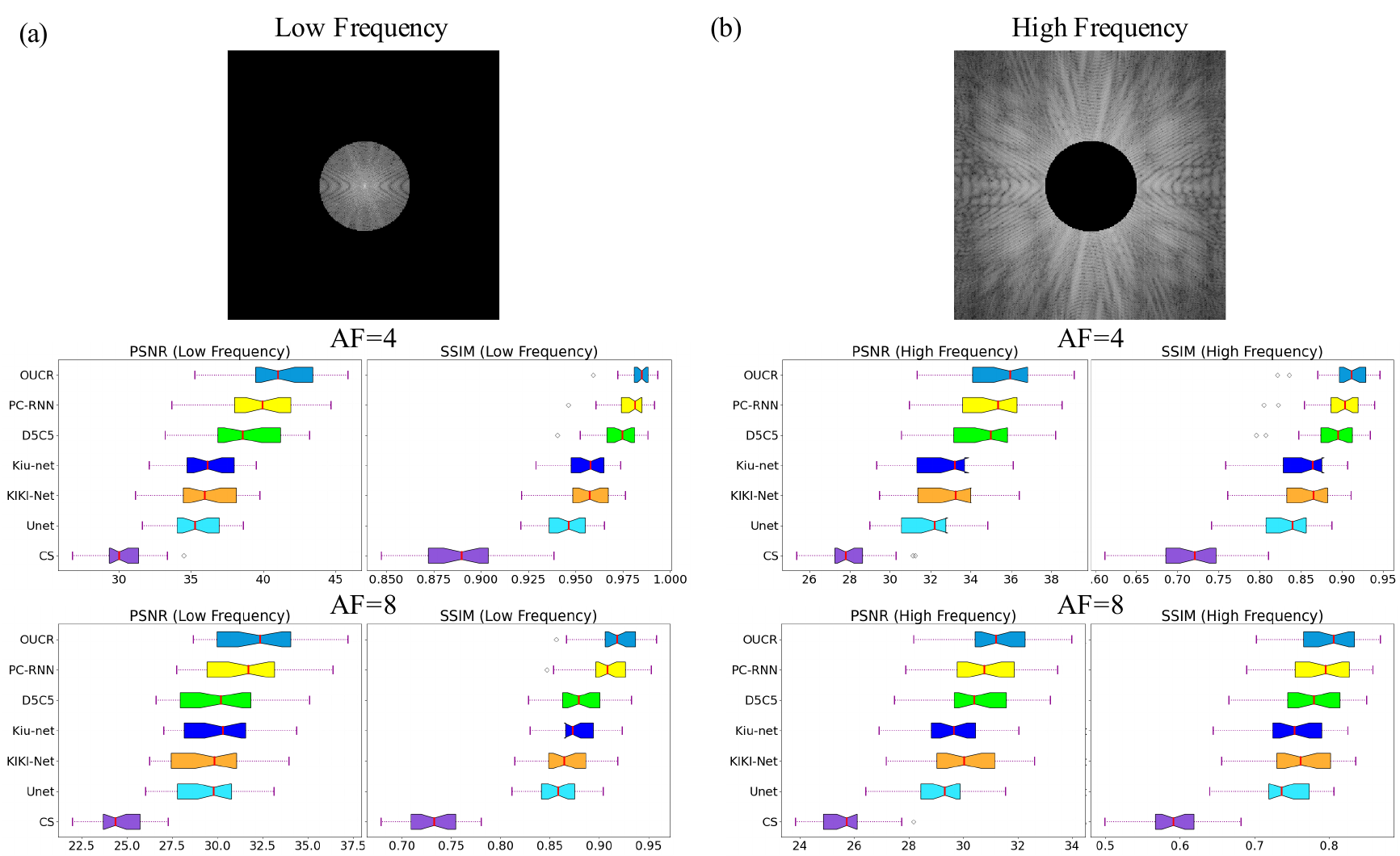}
	\vskip-12pt	\caption{K-space analysis on (a) low frequency and (b) high frequency. \emph{Top row}: the examples of masked k-space image. \emph{Bottom row}: the Boxplot of reconstruction performance by different methods on HPKS dataset.  \label{fig4}}
\end{figure}

In the k-space domain, the center frequencies determine the overall image contrast, brightness, and general shape. The peripheral area of k-space contains high spatial frequency information that controls edges, details, sharp transitions.~\cite{au26}. To further analyze the performance of different methods in low and high spatial frequency, we carry out the k-space analysis in Fig.~\ref{fig4}. We reconstruct an MR image from partially masked k-space and compare it with the reference image that is applied same mask on k-space, as shown in Fig.~\ref{fig4} top row. The reported PSNR and SSIM are presented by Boxplot in Fig.~\ref{fig4} bottom row. It can be seen that the proposed OUCR exhibits better reconstruction performance than other methods for both low and high frequency information.
		
\begin{table}[htp!]
	\centering
	\scriptsize
	\setlength{\tabcolsep}{25pt}
	\caption{Ablation study of designed modules in term of reconstruction quality on HPKS with AF=4.}\label{tab2}
\begin{tabular}{c|c|c}
	\hline
	Modules                       & PSNR           & SSIM            \\ \hline
	UC-CRNN                       & 37.87          & 0.9644          \\ \hline
	UC-CRNN + RM                  & 38.02          & 0.9658          \\ \hline
	OC-CRNN                       & 36.97          & 0.9564          \\ \hline
	OC-CRNN + RM                  & 37.34          & 0.9600          \\ \hline
	UC-CRNN + OC-CRNN             & 38.98          & 0.9719          \\ \hline
	UC-CRNN + OC-CRNN + RM (OUCR) & \textbf{39.33} & \textbf{0.9747} \\ \hline
\end{tabular}
\end{table}

\noindent {\bf{Ablation Study. }} We conduct a detailed ablation study to separately evaluate the effectiveness of using OC-CRNN, UC-CRNN, and RM in the proposed framework. The results are shown in Table~\ref{tab2}. We start with only using OC-CRNN and UC-CRNN. It can be noted that the performance of OC-CRNN is
lesser than UC-CRNN, since even though OC-CRNN captures the low-level features properly it does not capture most high-level features like UC-CRNN. Then, we show that adding RM with each individual module can improve the reconstruction quality. Finally, combing both CRNN networks with RM (proposed OUCR) results in the best performance. Fig.~\ref{fig5} illustrates the qualitative improvements after adding each major block, which is
consistent with the results reported in Table~\ref{tab2}. Moreover, we observe that increasing the number of CRNN iterations can further improve the performance of the proposed OUCR, but consequently leads to lower computational efficiency. Due to space constraint, an ablation study regarding CRNN iterations, k-space analysis on fastMRI dataset, and more visualizations are provided in the supplementary material. 

\section{Discussion and Conclusion}
We proposed a novel over-and-under complete convolutional RNN (OUCR) for MR image reconstruction. The purposed method leverages an overcomplete network to specifically capture low-level features, which are typically missed out in the other MR image reconstruction methods. Moreover, we incorporate an undercomplete CRNN, which results in an effective learning of low and high level information. 
% OUCR has less numbers of parameters comapred to previous methods.
The proposed method achieves better performance on two datasets and has less numbers of trainable parameters as compared to the CS and popular DL-based methods, including UNet~\cite{au24},  KIKI-Net~\cite{au12}, Kiu-net~\cite{au20}, D5C5~\cite{au13}, and PC-RNN~\cite{au15}. 
%For the fastMRI dataset, we only evaluate different methods on the fastMRI validation dataset, since the submission period of fastMRI leaderboard has ended. Therefore, our results are not directly comparable to the top leaderboard results on test dataset.
%Notably, the proposed OUCR model is fairly parameter-efficient compared to the other leaderboard models such as Adaptive-CS-Net (33M)~\cite{au29} and E2E-VN (30M)~\cite{au27,au28}. 
%This study demonstrates the potential of using overcomplete networks for MRI reconstruction.
This study demonstrates the potential of using overcomplete networks in MR image reconstruction task. 
%and provides the guidance of further extensions for MRI reconstruction.

\bibliographystyle{splncs04}
\bibliography{references}
\end{document}